\documentclass[aps, showpacs,preprintnumbers,amsmath,amssymb,superscriptaddress,twocolumn,nofootinbib]{revtex4-2}
\usepackage{amsmath,amsfonts,amssymb,graphics,graphicx,epsfig,color,times,indentfirst,layout,hyperref,multirow,bm}
\usepackage{hyperref, soul}
\hypersetup{linktocpage,colorlinks=true}
\usepackage{threeparttable}
\usepackage{makecell}
\usepackage{dcolumn}
\usepackage{bm}
\usepackage{color}
\usepackage{footnotebackref}
\def\mk{\mathbf{k}}
\newcommand{\corr}[1]{\textcolor{black}{#1}} % corrected
\newcommand{\comm}[1]{\textcolor{black}{#1}} % commented
\begin{document}
\title{Shift spin photocurrents in two-dimensional systems}

\author{Hsiu-Chuan Hsu}
\email{hcjhsu@nccu.edu.tw}\affiliation{Graduate Institute of Applied Physics, National Chengchi University, Taipei 11605, Taiwan}
\affiliation{Department of Computer Science, National Chengchi University, Taipei 11605, Taiwan}

\author{Tsung-Wei Chen}
\email{twchen@mail.nsysu.edu.tw}\affiliation{Department of
	Physics, National Sun Yat-sen University, Kaohsiung 80424, Taiwan}

\date{\today}

\begin{abstract}
	The generation of nonlinear spin photocurrents by circularly polarized light in two-dimensional systems is theoretically investigated
	by calculating the shift spin conductivities.
	In time-reversal symmetric systems, shift spin photocurrent can be generated under the irradiation of circularly polarized light , while the shift charge photoccurrent is forbidden by symmetry. We show that the $k$-cubic Rashba-Dresselhaus system, the $k$-cubic wurtzite system and Dirac surface states can support the shift spin photocurrent. By symmetry analysis, it is found that in the Rashba type spin-orbit coupled systems, mirror symmetry requires that the spin polarization and the moving direction of the spin photocurrent be parallel, which we name longitudinal shift spin photocurrent. 
		{The Dirac surface states with warping term exhibit mirror symmetry, similar to the Rashba type system, and support longitudinal shift spin photocurrent.}
	In contrast, in the Dresselhaus type spin-orbit coupled systems, the parity-mirror symmetry requires that the  spin polarization and the moving direction of the spin photocurrent be perpendicular, which we dub transverse shift spin photocurrent.
	Furthermore, we find that the shift spin photocurrent always vanishes in any $k$-linear spin-orbit coupled system unless the Zeeman coupling $\mu_z$ is turned on. We find that the splitting of degenerate energy bands due to Zeeman coupling $\mu_z$ causes the van Hove singularity. The resulting shift spin conductivity has a significant peak at optical frequency $\omega=2\mu_z/\hbar$. % When the Zeeman term is turned on, the shift spin photocurrent has non-zero longitudinal components for Rashba type spin-orbit coupling and non-zero transverse components for Dresselhaus type components. The effect of higher $k$ is also taken into account. 
	%The k-cubic Rashba always vanishes regardless of the Zeeman coupling. Nevertheless, the k-cubic Wurtzite system (Rashba type) keeps the non-zero longitudinal components of shift spin photocurrent in the absence of Zeeman coupling. 
	
\end{abstract}
 \maketitle
\section{introduction}
The bulk photovoltaic effect is the generation of DC current under the irradiation of light without applied bias; thus, %it plays an important role in green energy applications 
its  role in green energy applications has been actively investigated
\cite{Nagaosa2017,Cook2017,Pusch2023,Dai2023}. Moreover, the bias-free mechanism allows a noninvasive device fabrication of charge injection that reduces defects, leading to enhanced device quality \cite{Xu2021Pure}. 
Recent studies have demonstrated that the bulk photovoltaic effect is cosely related to the geometrical properties of Bloch states \cite{Morimoto2016a,Ahn2020,Ahn2022}, stretching the fundamental understanding of materials beyond energy bands \cite{Torma2023}. The bulk photovoltaic effect has been shown to originate from the Berry curvature \cite{deJuan2017}, quantum metric \cite{Gao2023} and Christoffel symbols \cite{Ahn2020,Hsu2023}.

The requirement for the bulk photovoltaic effects is the breaking of inversion symmetry \cite{Boyd}. Low-dimensional systems naturally lack the inversion symmetry and could support bulk photovoltaic effect \cite{Ganichev2014,Cook2017}. Typical examples include the two-dimensional electron gas in a semiconductor heterostructre and boundary modes of a bulk material.  
\corr{Furthermore, two types of spin-orbit coupling are often present in two-dimensional systems: Rashba spin-orbit coupling, which arises from broken structural inversion symmetry and can be induced by a built-in electric field, and Dresselhaus spin-orbit coupling, which results from inversion asymmetry in the bulk lattice \cite{Winkler,Ganichev2014}.} %The interplay between the spin-orbit coupling and quantum geometry rise to abundant bulk photovoltaic effect.  
The bulk photovoltaic effect is a second-order optical response and includes two types of mechanism: injection and shift photocurrents \cite{Aversa1995,Sipe2000}. {\footnote{In addition to current, the static magnetization can be generated by light in the second-order response, referred to as the nonlinear Edelstein effect \cite{Xu2021}.} The injection photocurrent is linearly proportional to the charge relaxation time, while the shift photocurrent is independent of the charge relaxation time. The shift photocurrent is described by the interband Berry connections of Bloch states and can be understood as the shift of wave packet upon interband transitions \cite{Nagaosa2017,Cook2017}. In time-reversal symmetric systems, the shift charge currents can be generated by linearly polarized light, but not by circularly polarized light \cite{Sipe2000,Ahn2020}.

In addition to the bulk photovoltaic effect for electron charges, the bulk photovoltaic effect for spins has attracted much attention \cite{Kim2017,Xu2021Pure,Fujimoto2024} for its potential application in spintronics.
It has been shown that spin photocurrent can be driven by the same mechanisms as the bulk photovoltaic effect, as derived by second-order response theory \cite{Kim2017,Xu2021Pure,Lihm2022}. Similar to the shift charge current, the shift spin photocurrent is related to the Berry connections, in addition to the spin current operator. 
 Due to the different selection rules for charge and spin, it has been shown that under certain symmetry constraints, it is possible to generate pure spin current without any charge current. %\corr{The influence of symmetry on pure spin photocurrent generation is discussed in spin-orbit coupled systems \cite{Xu2021Pure} and altermagnetic insulators which can support pure spin current regardless of SOC \cite{Dong2024,Pan2024}. } 
 \corr{Recently, pure spin photocurrent generation has been studied in altermagnetic insulators \cite{Dong2024,Pan2024}. 
 	The detailed symmetry analysis shows that when the altermagnetic insulators are irradiated by circularly polarized light, the shift spin photocurrent is zero in the absence of spin-orbit coupling, whereas the injection spin photocurrent can be nonzero regardless of spin-orbit coupling \cite{Dong2024}. }
 
 In time-reversal symmetric systems, the shift spin photocurrent can be generated under the circularly polarized light, while the shift charge current is forbidden.
However, the shift spin photocurrents under circularly polarized light have been overlooked until recently \cite{Xu2021Pure,Lihm2022}.  
%\comm{two paragraphs are combined.} 
Motivated by this issue, in this paper, we theoretically study the shift spin photocurrents generated by circularly polarized light 
%, here and after called shift spin photocurrent,
in perfect two-dimensional systems with spin-orbit interactions. In the absence of Zeeman coupling, we show that the shift spin photocurrents can be
generated only in some higher-$k$ systems (Wurtzite, $k$-cubic Dresselhaus and $k$-cubic Rashba-Dresselhaus systems) and Dirac surface states with preserved time-reversal symmetry.
In the $k$-linear systems, the shift spin photocurrent can only be generated in the presence of the Zeeman coupling. We also discuss how mirror symmetry constraints the
nonzero components of the shift spin conductivity tensors. The remainder of this paper is as follows. In Sec. \ref{sec:hamiltonian} and Sec. \ref{sec:shift}, the general two-band Hamiltonian and the shift spin conductivity are introduced, respectively. The results are given in the following sections. In Sec. \ref{sec:symm}, the symmetry analysis is given. In Sec. \ref{sec:iso}, the results for systems with isotropic dispersions are given. In Sec. \ref{sec:noniso}, the results for systems with non-isotropic dispersions are presented. Lastly, a conclusion is given in Sec. \ref{sec:disc}.

\section{effective Hamiltonian}\label{sec:hamiltonian}

The generic spin-orbit coupled Hamiltonian $H_0$ for the system with two energy bands can be written as
\begin{equation}\label{H1}
	H_0=\epsilon_{\mk}+\sigma_xd_x(\mk)+\sigma_yd_y(\mk)+\sigma_zd_z(\mk),
\end{equation}
where $\epsilon_k=\hbar^2k^2/2m$ is the kinetic energy, and $\sigma_x$, $\sigma_y$ and $\sigma_z$ are Pauli matrices \corr{representing spins of the charged particles}. We only consider two-dimensional systems, the momentum $\mathbf{k}$ has no $k_z$ component, i.e., $\mathbf{k}=(k_x,k_y)$. \corr{When an external magnetic field $B$ is applied along the $z$- axis,
	the Zeeman coupling is given by $\frac{1}{2}g\mu_B\sigma_zB$, where $g$ and $\mu_B$  represent g-factor and Bohr magneton, respectively. This coupling introduces an additional term to $d_z(\mathbf{k})$ in Eq. (\ref{H1}). In the following discussion, we define the Zeeman coupling as $\mu_z=\frac{1}{2}g\mu_BB$.}

We choose the eigenvectors of Eq. (\ref{H1}) as
\begin{equation}\label{H1-vector}
	\begin{split}
		&|{+\mk}\rangle=\frac{1}{\sqrt{2(1+\hat{d}_z)}}\left(\begin{array}{c}
			-\hat{d}_x+i\hat{d}_y\\
			1+\hat{d}_z\\
		\end{array}\right),\\
		&|{-\mathbf{k}}\rangle=\frac{1}{\sqrt{2(1+\hat{d}_z)}}\left(\begin{array}{c}
			1+\hat{d}_z\\
			\hat{d}_x+i\hat{d}_y\\
		\end{array}\right),
	\end{split}
\end{equation}
where $d=\sqrt{d_x^2+d_y^2+d_z^2}$ and $\hat{d}_x=d_x/d$, and so on. It is easy to show that Eq. (\ref{H1-vector}) satisfies the orthonormal property and completeness, i.e.,
\begin{equation}
	\sum_{n=\pm1}|{n\mk}\rangle\langle{n\mk}|=1,~~\langle{n\mk}|{m\mk}\rangle=\delta_{nm}.
\end{equation}
The corresponding eigen-energy is given by
\begin{equation}
	H_0|{n\mk}\rangle=E_{n\mk}|{n\mk}\rangle,~~n=\pm1
\end{equation}
with
\begin{equation}\label{eq:eigen}
	E_{n\mk}=\epsilon_{\mk}-nd(\mk).
\end{equation}
We note that the band index $n=+1$ corresponds to the lower energy $E_{+\mk}=\epsilon_{\mk}-d$ and $n=-1$ corresponds to the higher energy $E_{-\mk}=\epsilon_{\mk}+d$. The diagonal and off-diagonal matrix elements of spin operator in the basis Eq. (\ref{H1-vector}) are given in Appendix \ref{App:Spin}.

For typical spin-orbit coupled systems linearly in momentum $k$ are shown in Table \ref{Table-SOC-k1}, including Rashba~\cite{Rashba1960-1,Rashba1960-2,Rashba1960-3,Rashba1960-4}, Dresselhaus \cite{Dress1955-1,Dress1955-2,Dress1955-3,Dress1995-4}, Rashba-Dresselhaus, Weyl \cite{Weyl2015-1,Weyl2015-2}, persistent-spin-texture (PST) \cite{PST2018} and generic systems~\cite{Chen2013-1,Chen2013-2,Chen2013-3}. The PST system is a special case of linear Rashba-Dresselhaus system when $\beta_0=\alpha_0$. 
In Table \ref{Table-SOC-k1}, the functions $\Gamma_1(\phi)$, $\Gamma_2(\phi)$ and $\Gamma_3(\phi)$ are given by
\begin{equation}
	\begin{split}
		\Gamma_1(\phi)^2=&(\beta^2_{xx}+\beta^2_{yx})\cos^\phi+(\beta^2_{xy}+\beta^2_{yy})\sin^2\phi\\
		&+(\beta_{xx}\beta_{xy}+\beta_{yx}\beta_{yy})\sin(2\phi)
	\end{split}
\end{equation}
and
\begin{equation}\label{Gamma2}
	\begin{split}
		&\Gamma_2(\phi)=\sqrt{\alpha_0^2+\beta_0^2-2\alpha_0\beta_0\sin(2\phi)},\\
		&\Gamma_3(\phi)=\sqrt{2}\lambda\sqrt{1-\sin(2\phi)}.
	\end{split}
\end{equation}
The azimuthal angle $\phi$ is defined as $\tan\phi=k_y/k_x$. 
\begin{table}[]
	\centering
	\begin{tabular}{lll}
		\hline
		$k$-linear system & \multicolumn{1}{c}{$(d_x,d_y,d_z)$} & $d$ \\
		\hline\hline
		Generic &\makecell[l]{$(\beta_{xx}k_x+\beta_{xy}k_y$,\\$\beta_{yx}k_x+\beta_{yy}k_y,0)$} & $d=k\Gamma_1(\phi)$\\
		Rashba & $\alpha_0(k_y,-k_x,0)$ & $d=\alpha_0k$  \\
		Dresselhaus  & $\beta_0(-k_x,k_y,0)$  & $d=\beta_0k$  \\
		Rash.-Dress. & \makecell[l]{$(-\beta_0k_x+\alpha_0k_y$,\\$-\alpha_0k_x+\beta_0k_y,0)$} & $d=k\Gamma_2(\phi)$  \\
		Weyl & $\eta(k_x,k_y,0)$ & $d=\eta k$  \\
		PST & $\lambda(k_x-k_y,k_x-k_y,0)$ & $d=k\Gamma_3(\phi)$\\
		\hline
	\end{tabular}
	\caption{Five spin-orbit coupled systems with linear momentum in $k$.}\label{Table-SOC-k1}
\end{table}

Typical $k$-cubic systems are shown in Table \ref{Table-SOC-k3}, including Rashba ~\cite{Winkler2000-1,Winkler2000-2}, Dresselhuas ~\cite{Loss2005}, Rashba-Dresselhaus ~\cite{Mire2010}, and Wurtzite \cite{Wur1996-1,Wur1996-2} systems. In Table \ref{Table-SOC-k3}, the functions $\Gamma_2(\phi)$ for $k$-cubic Rashba-Dresselhaus system is the first line of Eq. (\ref{Gamma2}) by the replacement $\alpha_0\rightarrow\alpha$ and $\beta_0\rightarrow\beta$.

\begin{table}[]
	\centering
	\begin{tabular}{lll}
		\hline
		k-cubic system & \multicolumn{1}{c}{$(d_x,d_y,d_z)$} & $d$ \\
		\hline\hline
		Rashba& \makecell[l]{$\alpha(k_y(3k_x^2-k_y^2)$,\\$k_x(3k_y^2-k_x^2),0)$} & $d=\alpha k^3$  \\
		Dresselhaus & $-\beta k^2(k_x,k_y)$  & $d=\beta k^3$  \\
		Rash.-Dress.& \makecell[l]{$(\alpha k_y(3k_x^2-k_y^2)-\beta k^2k_x,$\\$\alpha k_x(3k_y^2-k_x^2)-\beta k^2k_y,0)$} & $d=k^3\Gamma_2(\phi)$  \\
		Wurtzite & $(\alpha'+\beta' k^2)(k_y,-k_x,0)$ & $d=\alpha' k+\beta'k^3$  \\
		\hline
	\end{tabular}
	\caption{Four spin-orbit coupled systems with $k^3$.}\label{Table-SOC-k3}
\end{table}

\section{Shift spin conductivity}\label{sec:shift}
In this section, we calculate the shift spin conductivity for the system Hamiltonian Eq. (\ref{H1}). The shift spin (charge) conductivity is written as \cite{Lihm2022}
\begin{equation}\label{ShiftSpin-1}
	\sigma^{Ic;ab}=\frac{-i\pi e^3}{\hbar^2}\int_{\mk}\sum_{nm}(f_n-f_m)M^{Ic;ab}_{mn}\delta(\omega_{mn}-\omega),
\end{equation}
where $f_n$ represents Fermi-Dirac distribution and $\omega_{mn}=(E_m-E_n)/\hbar$. The
superscript $I=0$ represents the shift charge current and $I=x,y,z$ means spin polarization. We focus on the shift spin photocurrent induced by the circular light ($a=x,b=y$), and thus, we consider the imaginary part of Eq. (\ref{ShiftSpin-1}),
\begin{equation}\label{ShiftSpin-3}
	Im[\sigma^{Ic;ab}]=\frac{-\pi e^3}{\hbar^2}\int_{\mk}\sum_{nm}(f_n-f_m)Re[M^{Ic;ab}_{mn}]\delta(\omega_{mn}-\omega).
\end{equation}
The matrix element $M_{mn}^{Ic;ab}$ is given by
\begin{equation}\label{ShiftSpin-2}
	M_{mn}^{Ic;ab}=\frac{v_{nm}^b}{\omega_{mn}^2}K^{ca}_{mn}-\frac{v_{mn}^a}{\omega^2_{nm}}K^{cb}_{nm},
\end{equation}
where $K^{ca}_{mn}$ and $K^{cb}_{nm}$ are given by
\begin{equation}
	\begin{split}
		&K^{ca}_{mn}=d_{mn}^{ca}-L^{ca}_{mn},\\
		&K^{cb}_{nm}=d_{nm}^{cb}-L^{cb}_{nm},
	\end{split}
\end{equation}
and matrix element $d_{\cdot\cdot}^{\cdot\cdot}$ is %the spin-velocity derivative of the velocity operator~\cite{Lihm2022}
\begin{equation}\label{eq:d-virt}
	\begin{split}
		&d_{mn}^{ca}=J_{mn}^{Ica}+\sum_{p\neq m,n}\left(\frac{J^{Ic}_{mp}v^a_{pn}}{\omega_{mp}}-\frac{v^a_{mp}J^{Ic}_{pn}}{\omega_{pn}}\right),\\
		&d_{nm}^{cb}=J_{\corr{nm}}^{Icb}+\sum_{p\neq m,n}\left(\frac{J^{Ic}_{np}v^b_{pm}}{\omega_{np}}-\frac{v^b_{np}J^{Ic}_{pm}}{\omega_{pm}}\right),\\
	\end{split}
\end{equation}
where
\begin{equation}
	\corr{J^{Ica}}=\frac{1}{2}\left\{s_I,\frac{\partial^2H_0}{\partial k_a\partial k_c}\right\},~~J^{Ic}=\frac{1}{2}\{s_I,v^c\},
\end{equation}
and $s_I=\frac{1}{2}\sigma_I$ and $\sigma_I$ stands for the Pauli matrix. The velocity operator $v^c$ is given by $v^c=\frac{\partial(H_0/\hbar)}{\partial k_c}$. We note that the last term in the matrix element $d_{\cdot\cdot}^{\cdot\cdot}$ is the virtual transition. The dummy index $p$ in Eq. (\ref{eq:d-virt}) stands for the summation over all bands except for the band indices $m,n$. 
\corr{%For effective Hamiltonians with a finite number of bands, 
	The dominant contributions to the summation come from bands near the Fermi level within the photon energy \cite{Ventura2017,Xu2021Pure}. Thus, the error for effective Hamiltonians with finite number of bands is negligible in the low-frequency regime. A truncation-error-free scheme for shift charge current, particularly useful in {\it ab initio} calculations, has been developed in Ref. \cite{Azpiroz2018}. }
 All the spin-orbit coupled systems considered in Table I and Table II describe spin-$1/2$ particles, except for the $k$-cubic Rashba and Dresselhaus Hamiltonians, which describe the heavy hole bands with \comm{total angular momentum} $3/2$ and \comm{the} z components of $\pm3/2$ \cite{Winkler2000-1, Winkler2000-2}. Thus,  for $k$-cubic Hamiltonians $s_I$ is replaced by $3s_I$. \comm{For convenience, the transport of the total angular momentum in cubic Hamiltonians is referred to as spin current, as used in the literature (e.g., \cite{Schlimann2005,Nakamura2012}) and discussed in App. \ref{app:spinop}.}}
On the other hand, the matrix element $L^{\cdot\cdot}_{\cdot\cdot}$ 
\corr{ is given by }
\begin{equation}
	\begin{split}
		&L_{mn}^{ca}=\frac{1}{\omega_{mn}}(J^{Ic}_{mn}\Delta^a_{mn}+v^{a}_{mn}\Delta^{Ic}_{mn}),\\
		&L_{nm}^{cb}=\frac{1}{\omega_{nm}}(J^{Ic}_{nm}\Delta^b_{nm}+v^{b}_{nm}\Delta^{Ic}_{nm}),
	\end{split}
\end{equation}
where $\Delta_{mn}^{a}=v^a_{mm}-v^a_{nn}$ and \corr{$\Delta_{mn}^{Ic}=J^{Ic}_{mm}-J^{Ic}_{nn}$}. We note that $\omega_{mn}^2=\omega_{nm}^2$ and Eq. (\ref{ShiftSpin-2}) can be written as
\begin{equation}
	\begin{split}
		&M_{mn}^{Ic;ab}\\
		&=\frac{1}{\omega^2_{nm}}\left[(v^b_{nm}d^{ca}_{mn}-v^a_{mn}d^{cb}_{nm})-(v^b_{nm}L^{ca}_{mn}-v^a_{mn}L^{cb}_{nm})\right]\\
		&=\frac{1}{\omega^2_{nm}}(M_d-M_L),
	\end{split}
\end{equation}
where $M_d$ and $M_L$ are defined as
\begin{equation}\label{Eq:MdMl}
	\begin{split}
		M_d=(v^b_{nm}d^{ca}_{mn}-v^a_{mn}d^{cb}_{nm}),\\
		M_L=(v^b_{nm}L^{ca}_{mn}-v^a_{mn}L^{cb}_{nm}).
	\end{split}
\end{equation}

For the present case, we consider a two-band model [see Eq. (\ref{H1})] and hence there is no virtual band transition. This implies that the matrix element $d^{\cdot\cdot}_{\cdot\cdot}$ is given by
\begin{equation}
	\begin{split}
		&d_{mn}^{ca}=J_{mn}^{Ica},\\
		&d_{nm}^{cb}=J_{\corr{nm}}^{Icb}.\\
	\end{split}
\end{equation}
Furthermore, the term $\Delta^{Ic}_{mn}$ will be exactly canceled in deriving $M^{Ic;ab}$, we can neglect it in the derivation. That is, we have
\begin{equation}
	\begin{split}
		&L_{mn}^{ca}=\frac{1}{\omega_{mn}}J^{Ic}_{mn}\Delta^a_{mn},\\
		&L_{nm}^{cb}=\frac{1}{\omega_{nm}}J^{Ic}_{nm}\Delta^b_{nm}.
	\end{split}
\end{equation}

Because $\omega>0$, $\omega_{mn}$ in the Dirac-delta function must be positive. This implies that $m=-1$ and $n=+1$. After straightforward calculations, we have
\begin{equation}\label{Eq-Md}
	M_d=v^b_{1,-1}J^{Ica}_{-1,1}-v^a_{-1,1}J^{Icb}_{1,-1}
\end{equation}
and
\begin{equation}\label{Eq-ML}
	M_L=\frac{1}{d}\left[v^b_{1,-1}J^{Ic}_{-1,1}\frac{\partial d}{\partial k_a}-v^a_{-1,1}J^{Ic}_{1,-1}\frac{\partial d}{\partial k_b}\right],
\end{equation}
where we have used $E_{+\mk}-E_{-\mk}=-2d$ and $E_{-\mk}-E_{+\mk}=2d$. The resulting matrix element $M^{Ic,ab}_{mn}$ is given by
\begin{equation}\label{Eq-M-Icab}
	M_{-1,1}^{Ic,ab}=\frac{\hbar^2}{4d^2}(M_d-M_L).
\end{equation}
If we use the replacement $a\leftrightarrow b$, which means the change in the helicity of the circular polarized light, then by using Eqs. (\ref{Eq-Md}), (\ref{Eq-ML}) and (\ref{Eq-M-Icab}), we have
\begin{equation}
	Re[M^{Ic,ba}_{-1,1}]=-Re[M^{Ic,ab}_{-1,1}],
\end{equation}
which only gives an overall sign. In the following calculations, we only calculate the six matrix elements of $Re[M^{Ic,xy}_{-1,1}]$, i.e., $I=x,y,z$ and $c=x,y$. The six matrix elements for the two-band model are given in Appendix \ref{App:Matrix-dz}. For the two-dimensional cases, the spin lies on the two-dimensional plane and the spin $z$ component only couples to the Zeeman interaction $\mu_z$. Accounting for the case $d_z=\mu_z$, Eq. (\ref{App:M-Icab})  can be written as the following form. For in-plane spin $I=x,y$, we have
\begin{widetext}
	\begin{equation}\label{Md-ML-1}
		Re[M^{Ic,xy}_{-1,1}]=N\left\{-sgn[\hat{e}_1]\frac{\mu_z^2}{d^2}\left[\frac{\partial d_I}{\partial k_{\hat{e}_1}}\left(1-\frac{k_c}{d}\frac{\partial d}{\partial k_c}\right)+\frac{\partial d_I}{\partial k_c}\frac{k_c}{d}\frac{\partial d}{\partial k_{\hat{e}_1}}\right]+sgn[\hat{e_2}]G\frac{\hat{d}_{\hat{e}_2}}{d}k_c\left(1-\frac{k}{d}\frac{\partial d}{\partial k}\right)\right\}.
	\end{equation}
\end{widetext}
For out-of-plane spin $I=z$ and $c=x,y$, we have
\begin{equation}\label{Md-ML-2}
	Re[M^{zc,xy}_{-1,1}]=N sgn[\hat{e}_1]\frac{\mu_z}{d}\frac{\partial d}{\partial k_{\hat{e}_1}},
\end{equation}
where we use the notation
\begin{equation}
	\begin{split}
		&\hat{e}_1=\hat{e}_c\times\hat{e}_z,\\
		&\hat{e}_2=\hat{e}_I\times\hat{e}_z,
	\end{split}
\end{equation}
and we define $sgn[\hat{e}_x\times\hat{e}_z]=sgn[-\hat{e}_y]=-1$ and $sgn[\hat{e}_y\times\hat{e}_z]=sgn[+\hat{e}_x]=+1$. The term $G$ and $N$ are given in Eq. (\ref{App:M-Icab2}), (\ref{App:M-Icab3}) and (\ref{App:M-Icab4}). The in-plane component Eq. (\ref{Md-ML-1}) is composed of two terms, one is proportional to the square of Zeeman coupling $\mu_z^2$ and the other term is independent of Zeeman coupling. The out-of-plane component Eq. (\ref{Md-ML-2}) is linearly proportional to Zeeman coupling. When we turn off the Zeeman coupling, the out-of-plane component always vanishes, but the in-plane components may be non-zero. In the following sections, for the convenience of discussion, the matrix elements  $Re[M^{xx,xy}_{-1,1}]$ and $Re[M^{yy,xy}_{-1,1}]$ (i.e., $I=c$) are called longitudinal components, and the matrix elements  $Re[M^{xy,xy}_{-1,1}]$ and $Re[M^{yx,xy}_{-1,1}]$ (i.e., $I\neq c$) are called transverse components. The matrix elements $Re[M^{zx,xy}_{-1,1}]$ and $Re[M^{zy,xy}_{-1,1}]$ (i.e., $I=z$) are called out-of-plane components.

\section{Symmetry analysis}\label{sec:symm}
The two-dimensional systems studied in this work preserve mirror or parity-mirror symmetry. We discuss the symmetry constraints on shift spin conductivity imposed by the symmetry for the Rashba and Dresselhaus type Hamiltonians that commonly arise in two-dimensional systems. For simplicity, we take the $k$-linear Rashba and Dresselhaus spin-orbit coupling (SOC) as an example
\begin{eqnarray}
	h_R&=&\alpha_0(k_y\sigma_x-k_x\sigma_y),\\
	h_D&=&\beta_0(k_x\sigma_x-k_y\sigma_y).
\end{eqnarray}
The $k$-cubic counterparts share the same symmetry and the constraints.  
Under inversion symmetry, $k_x \rightarrow -k_x, k_y\rightarrow -k_y$, while the spins do not change sign. Thus, both SOC types break inversion symmetry. 

Under mirror symmetry $M_x$, where $k_x \rightarrow -k_x, k_y\rightarrow k_y$ and $\sigma_x \rightarrow \sigma_x, \sigma_{y,z}\rightarrow -\sigma_{y,z}$, it can be shown that $h_R$ is invariant. In contrast, $h_D$ is odd under $M_x$. 
The similar operation can be applied to mirror symmetry $M_y$. It is shown that $h_R$ is invariant, whereas $h_D$ is odd. The symmetry of the two common types of SOC in two-dimensional systems are summarized in Tab. \ref{tab:socsym}.

\begin{table}[]
	\begin{tabular}{|l|c|c|c|c|}
		\hline
		& $P$ & $M_x$ & $M_y$ & $C_2$ \\ \hline
		\begin{tabular}[c]{@{}l@{}}Rashba type\\ (linear, cubic, Wurtzite)\end{tabular} & -   & +     & +   &+   \\ \hline
		\begin{tabular}[c]{@{}l@{}}Dresselhaus type\\ (linear, cubic)\end{tabular}      & -   & -     & - &+     \\ \hline
	\end{tabular}
	\caption{The symmetry properties of the SOC Hamiltonians under inversion ($P$), mirror symmetry ($M_x$, $M_y$) and rotation by $\pi$ ($C_2$). The $\pm$ sign denotes the evenness, oddness of the Hamiltonian under the symmetry operation.}
	\label{tab:socsym}
\end{table}

The mirror symmetry constraints nonvanishing components of the shift spin conductivity tensors. In this work, we consider the response to circularly polarized light normally incident on the two-dimensional plane. In this scenario, the spatial symmetry properties of the matrix element $M_{mn}^{Ic;ab}$ in the shift spin conductivity are the same as those of the product of the spin current ($J_{mn}^{I,c}$) and the velocity operators ($v_{mn}^xv_{nm}^y$), as shown from Eq. (\ref{Eq:MdMl}). If the product is an odd function in the Brillouin zone, it leads to vanishing shift spin conductivities. 
$h_R$ is $M_{x}$ and $M_y$ symmetric, such that $v_{mn}^xv_{nm}^y$ is an odd function under $M_x$ or $M_y$. $J_{mn}^{I,c}$, given by the product of the spin and the velocity operator, also has to be an odd function to have nonzero responses. The mirror symmetry operation $M_d,d=x,y$, transforms the spin  matrix elements  ${s}^I_{mn}(\mathbf{k})$ to $(-1)^{\delta_{d,I}+1}s^I_{mn}(M_d\mathbf{k})$ and the velocity matrix elements 
${v}^c_{mn}(\mathbf{k})$ to $(-1)^{\delta_{d,c}}v^c_{mn}(M_d\mathbf{k})$ \cite{Ahn2020,Xu2021Pure,Lihm2022}.  Thus,  only when the spin polarization is parallel to $c$, there is nonzero shift spin photocurrents. We dub this response as the longitudinal shift spin photocurrent. 

$h_D$ is even under the parity-mirror symmetry operation $PM_x$ and $PM_y$, such that  $v_{mn}^xv_{nm}^y$ is an odd function. $J_{mn}^{I,c}$ has to be an odd function under this parity-mirror operation to have nonzero responses. The parity-mirror operation
transforms the spin  matrix elements  ${s}^I_{mn}(\mathbf{k})$ to $(-1)^{\delta_{d,I}+1}s^I_{mn}(-M_d\mathbf{k})$ and the velocity matrix elements 
${v}^c_{mn}(\mathbf{k})$ to $(-1)^{\delta_{d,c}+1}v^c_{mn}(-M_d\mathbf{k})$ \cite{Ahn2020,Xu2021Pure,Lihm2022}. 
Thus, only when the spin polarization is perpendicular to $c$, there is nonzero shift spin photocurrents. We call this response as the transverse shift spin photocurrent. Furthermore, Dresselhaus SOC is $M_yM_x$ symmetric under which $v_{mn}^xv_{nm}^y$ is even and $J_{mn}^{z,x}, J_{mn}^{z,y}$ are odd. Consequently, the spin current with out-of-plane spin polarization is zero. The symmetry analysis shows that for Rashba type spin-orbit coupled systems, the nonzero components $Ic$ of the shift spin photocurrents are $xx$, $yy$. For Dresselhaus type, the nonzero components are $xy$, $yx$.

\section{Isotropic energy dispersion}\label{sec:iso}
For the system with isotropic energy dispersion, we mean that the band gap $d(k_x,k_y)$ has the following form
\begin{equation}\label{Iso-d}
	d=k^q\gamma,
\end{equation}
where $q=1,2,3\cdots$ and $\gamma$ is only a function of spin-orbit couplings and is independent of the azimuthal angle $\phi=tan^{-1}(k_y/k_x)$. For the present cases under consideration, we have $d=\alpha_0 k$ ($k$-linear Rashba coupling), $d=\beta_0 k$ ($k$-linear Dresselhaus system), $d=\alpha k^3$ ($k$-cubic Rashba system), $d=\beta k^3$ ($k$-cubic Dresselhaus system) and $d=\alpha k+\beta k^3$ (Wurtize system). In the presence of Zeeman coupling, the dispersion becomes
\begin{equation}
	d=\sqrt{(k^q\gamma)^2+\mu_z^2}.
\end{equation}
When $\mu_z=0$, Eq. (\ref{Md-ML-1}) becomes
\begin{equation}
	Re[M^{Ic,xy}_{-1,1}]=N\cdot sgn[\corr{\hat{e}_2}]G\frac{\hat{d}_{\corr{\hat{e}_2}}}{d}k_c\left(1-\frac{k}{d}\frac{\partial d}{\partial k}\right).
\end{equation}
Furthermore, we have
\begin{equation}\label{Md-ML-3}
	\left(1-\frac{k}{d}\frac{\partial d}{\partial k}\right)=(1-q),~~\mu_z=0.
\end{equation}
For $k$-linear system $q=1$, we always have
\begin{equation}
	Re[M^{Ic,xy}_{-1,1}]=0.
\end{equation}
Namely, for $k$-linear system without Zeeman coupling, the shift spin photocurrent cannot be generated by the circular polarized light. For $k$-cubic systems, $\left(1-\frac{k}{d}\frac{\partial d}{\partial k}\right)=(1-3)=-2$, it is possible to have non-zero shift spin photocurrent, but it depends on the integration of $Re[M^{Ic,xy}]$, which should be confirmed by numerical calculations and symmetry constraints shown in Sec. \ref{sec:symm}.

When the Zeeman term is taken into account, the energy dispersion is $d^2=(k^q\gamma)^2+\mu_z^2$, and thus, in general $\left(1-\frac{k}{d}\frac{\partial d}{\partial k}\right)$ is not equal to zero even in $k$-linear systems.
For pure $k$-linear Rashba system, by using Eq. (\ref{App:M-Icab}), we have
\begin{equation}
	\begin{split}
		&Re[M^{yy,xy}_{-1,1}]=Re[M^{xx,xy}_{-1,1}]=N\frac{\alpha_0\mu_z^2}{d^2},\\
		&Re[M^{xy,xy}_{-1,1}]=Re[M^{yx,xy}_{-1,1}]=0,\\
		&Re[M^{zx,xy}_{-1,1}]=-N\frac{\mu_z}{d^2}\alpha^2_0k_y,\\
		&Re[M^{zy,xy}_{-1,1}]=N\frac{\mu_z}{d^2}\alpha^2_0k_x.\\
	\end{split}
\end{equation}
Because the term $N$ and $d$ are independent of $\phi$, we observe that after integration of $\phi$, only two terms survive, which are $\sigma^{xx,xy}$ and $\sigma^{yy,xy}$. For a pure $k$-linear Dresselhaus system, by using Eq. (\ref{App:M-Icab}), we have
\begin{equation}
	\begin{split}
		&Re[M^{yy,xy}_{-1,1}]=Re[M^{xx,xy}_{-1,1}]=0,\\
		&Re[M^{xy,xy}_{-1,1}]=Re[M^{yx,xy}_{-1,1}]=-N\frac{\beta_0\mu_z^2}{d^2},\\
		&Re[M^{zx,xy}_{-1,1}]=-N\frac{\mu_z}{d^2}\beta^2_0k_y,\\
		&Re[M^{zy,xy}_{-1,1}]=N\frac{\mu_z}{d^2}\beta^2_0k_x.
	\end{split}
\end{equation}
Because the terms $N$ and $d$ are independent of $\phi$, we observe that after integration of $\phi$, only two terms survive, which are $\sigma^{xy,xy}$ and $\sigma^{yx,xy}$. These results are consistent with the symmetry consideration in Sec. \ref{sec:symm}. Furthermore, the nonvanishing shift spin conductivities are proportional to the strength of spin-orbit coupling. These results are listed in Table \ref{Table-Shift-RD}, where all the shift spin conductivity vanishes when $\mu_z=0$.

\begin{table}
	\centering
	\begin{tabular}{|c|c|cccccc|}
		\hline
		$k$-linear systems & $d^2$ & xx & xy & yx & yy & zx & zy  \\
		\hline\hline
		Rashba &$\alpha_0^2 k^2+\mu_z^2$& \checkmark & 0 & 0 & \checkmark & 0 & 0  \\
		Dresselhaus&$\beta_0^2k^2+\mu_z^2$  & 0 & \checkmark & \checkmark & 0 & 0 &0   \\
		Weyl&$\eta^2k^2+\mu_z^2$  & 0 & \checkmark & \checkmark & 0 & 0 &0\\
		\hline
	\end{tabular}
	\caption{The shift spin conductivity for k-linear systems: pure Rashba (R), pure Dresselhaus (D) and Weyl systems. {The nonvanishing components are labeled by $\checkmark$, whereas the vanishing components are listed as $0$.} Note that when the Zeeman coupling is turned off ($\mu_z=0$), all components vanish for all systems. These results are consistent with the numerical results.}\label{Table-Shift-RD}
\end{table}

\begin{table}
	\centering
	\begin{tabular}{|c|c|cccccc|}
		\hline
		$k$-cubic systems & $d^2$ & xx & xy & yx & yy & zx & zy  \\
		\hline\hline
		Rashba&$\alpha^2 k^6+\mu_z^2$& 0 & 0 & 0 & 0 & 0 & 0  \\
		Dresselhaus&$\beta^2k^6+\mu_z^2$& 0 & \checkmark & \checkmark & 0 & 0 &0   \\
		Wurtzite&$(\alpha'+\beta'k^2)^2k^2+\mu_z^2$& \checkmark & 0 & 0& \checkmark& 0& 0\\
		\hline\hline
		Rashba&$\alpha^2 k^6$& 0 & 0 & 0 & 0 & 0 & 0  \\
		Dresselhaus&$\beta^2k^6$& 0 & \checkmark & \checkmark & 0 & 0 &0   \\
		Wurtzite&$(\alpha'+\beta'k^2)^2k^2$& \checkmark & 0 & 0& \checkmark& 0& 0\\
		\hline
	\end{tabular}
	\caption{The shift spin conductivity for k-cubic systems: pure Rashba (R), pure Dresselhaus (D), and Wurtzite (W) systems. The upper (lower) row corresponds to the pure Rashba (pure Dresselhaus) system. {The nonvanishing components are labeled by $\checkmark$, whereas the vanishing components are listed as $0$. These results are consistent with the numerical results.}}\label{Table-Shift-k3}
\end{table}

The shift spin conductivities for $k$-cubic systems are listed in Table \ref{Table-Shift-k3}, where all the non-zero components are still non-zero when $\mu_z=0$ except for the $k$-cubic Rashba system. We note that in $k$-cubic Rashba system, all the components vanish after integration over $\phi$.

It is interesting to note that the wurtzite system has dispersion similar to that of the pure $k$-linear Rashba system. Comparing the two  systems, we have $d_x=f(k)k_y$ and $d_y=-f(k)k_x$ and $f(k)=\alpha$ for the pure $k$-linear Rashba system and $f(k)=\alpha'+\beta' k^2$ for the Wurtzite system. The presence of higher-order $k$ dependence, $\beta'$, leads to the result that $Re[M^{Ic,xy}_{-1,1}]$ is proportional to $\beta'$, and in fact we have
\begin{equation}\label{k3-W-beta}
	\begin{split}
		&Re[M^{yy,xy}_{-1,1}]=-\frac{2Gk^2\beta'\sin^2\phi}{(\alpha'+\beta'k^2)^2},\\
		&Re[M^{xx,xy}_{-1,1}]=-\frac{2Gk^2\beta'\cos^2\phi}{(\alpha'+\beta'k^2)^2},\\
		&Re[M^{xy,xy}_{-1,1}]=Re[M^{yx,xy}_{-1,1}]=-\frac{Gk^2\beta'\sin(2\phi)}{(\alpha'+\beta'k^2)^2},\\
	\end{split}
\end{equation}
When $\beta'$ vanishes, Eq. (\ref{k3-W-beta}) goes back to the result of the pure Rashba system. Because of the higher-order $k$-dependence in $\beta'$, we find that the longitudinal shift spin photocurrent survives in the absence of Zeeman coupling.

\corr{Before closing this section, we discuss the effect of the isotropic higher-order $k$ terms, in the form of $\vec{d'}=k^{q-1}\vec{d}$. The introduction of such terms does not change the symmetry of the Hamiltonian or the isotropy of the dispersion. \comm{The spin textures are not changed, either. As shown by Eq. \ref{Pauli-1}, the spin polarization is parallel to the normalized $\vec{d}$-vector.} Notably, we find that the addition of the isotropic higher-order $k$ terms to the $k$-linear Rashba and Dresselhaus systems could lead to nonzero shift spin photocurrent.}

\corr{Consider the Rashba-type Hamiltonian with isotropic  higher-order $k$. The $d_x$ and $d_y$ components in the Hamiltonian become }
\begin{equation}
	d_x=\alpha k^{q-1}k_y,~~d_y=-\alpha k^{q-1}k_x,
\end{equation}
and by using Eq. (\ref{App:Def-G}), we have $G=\alpha^2k^{2(q-1)}$. Inserting these equation into Eq. (\ref{App:M-Icab}), after integration of $\phi$, we have
\begin{equation}\label{App:R-type}
	\begin{split}
		Re[M^{xx,xy}_{-1,1}]&=Re[M^{yy,xy}]\\
		&=\frac{\alpha\pi}{d^2}k^{q-1}\left[\mu_z^2(q+1)-\alpha^2k^{2q}(q-1)\right],\\
		Re[M^{xy,xy}_{-1,1}]&=Re[M^{yx,xy}_{-1,1}]=0.\\
	\end{split}
\end{equation} 
When $q=1$ (pure Rashba), the shift spin photocurrent can be generated in the presence of Zeeman coupling. When $q\neq1$, the shift spin photocurrent would be non-zero in the absence of Zeeman coupling. The term $q=3$ corresponds to one part of Wurtzite system, and thus, in Wurtzite system there are non-zero shift spin photocurrent in the absence of Zeeman coupling. Furthermore, only longitudinal components ($xx$ and $yy$) survive, the transverse terms ($xy$ and $yx$) always vanish. 
We note that the above discussion does not apply to the $k$-cubic Rashba system with $d^R_x=\alpha k_y(3k_x^2-k_y^2)$ and $d^R_y=\alpha k_x(3k_y^2-k_x^2)$, because it does not relate to $k$-linear Rashba system by a factor of isotropic higher-order $k$ terms. \comm{The difference between the $k$-linear and $k$-cubic Rashba Hamiltonians can be manifested in the spin textures on Fermi contours, as shown in Fig. \ref{fig:spintext} in App. \ref{app:numer}.}

Consider the Dresselhaus-type Hamiltonian with higher-order $k$,
\begin{equation}
	d_x=\beta k^{q-1}k_x,~~d_y=\pm\beta k^{q-1}k_y,
\end{equation}
where the plus sign corresponds to the $k$-cubic type Dresselhaus system ($G=\beta^2k^{2q-2}$), and the minus sign corresponds to the $k$-linear type Dresselhaus system ($G=-\beta^2k^{2q-2}$). By using Eq. (\ref{App:M-Icab}) and after integration of $\phi$, we have
\begin{equation}\label{App:D-type}
	\begin{split}
		Re[M^{xx,xy}_{-1,1}]&=Re[M^{yy,xy}]=0,\\
		Re[M^{xy,xy}_{-1,1}]&=-Re[M^{yx,xy}]\\
		&=\frac{\beta\pi}{d^2}k^{q-1}\left[k^{2q}\beta^2(q-1)-\mu_z^2(q+1)\right].
	\end{split}
\end{equation} 
Unlike the Rashba-type Hamiltonian, the Dresselhaus type Hamiltonian has the non-zero transverse components and the longitudinal components vanish. Moreover, similar to the Rashba-type Hamiltonian, the higher-order term is the key to having nonzero shift spin photocurrent in the absence of Zeeman coupling. If $q=1$, i.e., $k$-linear Dresselhaus, the shift spin photocurrent always vanishes without Zeeman coupling.

%On the other hand, the measurement of the shift spin photocurrent enables us to detect the existence of the higher order $k$ terms. Consider the Rashba type Hamiltonian with $d_x=\alpha k^{q-1}k_y$ and $d_y=-\alpha k^{q-1}k_x$. In the presence of Zeeman coupling, its longitudinal components are non-zero [see Eq. (\ref{App:R-type})] in this case. The Dirac delta function gives $\omega=2d/\hbar$ and we can obtain the momentum $k$ as a function of the circular polarized light $\omega$ denoted as $k(\omega)$. When applying frequency such that the non-zero component $Re[M^{xx,xy}_{-1,1}]$ and $Re[M^{yy,xy}_{-1,1}]$ vanish, we have
%\begin{equation}
%	\left\{\begin{array}{c}
%		\displaystyle k^{2q}\alpha^2(q-1)-\mu_z^2(q+1)=0\\
%		\displaystyle \frac{2}{\hbar}\sqrt{\alpha^2k^{2q}+\mu_z^2}=\omega\\
%	\end{array}\right.
%\end{equation}
%and the resulting $q$ is given by
%\begin{equation}\label{Eq-q-exp}
%	q=\frac{1}{1-\chi},~~\chi=\frac{8\mu_z^2}{\hbar^2\omega^2}
%\end{equation}
%For Dresselhaus-type systems, we have the same result for non-zero transverse components [see Eq. (\ref{App:D-type})]. Therefore, by using Eq. (\ref{Eq-q-exp}), we can determine the higher order $q$-value of the system. In other words, the frequency ceasing the non-zero shift current would be
%\begin{equation}
%	\omega=\frac{2\mu_z}{\hbar}\sqrt{\frac{2q}{q-1}}
%\end{equation}
%For the present system, if the system is a perfect k-linear system ($q=1$), we have $\omega=\infty$, and if the system is a perfect k-cubic system ($q=3$), we have $\omega=2\sqrt{3}\mu_z/\hbar$.

\section{Non-isotropic energy dispersion}\label{sec:noniso}
For the system with non-isotropic energy dispersion, we mean its band gap $d$ is written as
\begin{equation}\label{NonIso-d}
	d=k^q\Gamma(\phi),
\end{equation}
where $\Gamma(\phi)$ is a function of azimuthal angle $\phi$ and spin-orbit couplings. The typical systems under consideration are the $k$-linear Rashba-Dresselhaus system $(d=k\Gamma_2(\phi))$, $k$-cubic Rashba-Dresselhaus system $(d=k^3\Gamma_2(\phi))$, and the Dirac surface state with the warping term.

\subsection{The $k$-linear and cubic Rashba-Dresselhaus systems}
\begin{figure}
	\includegraphics[width=0.45\textwidth]{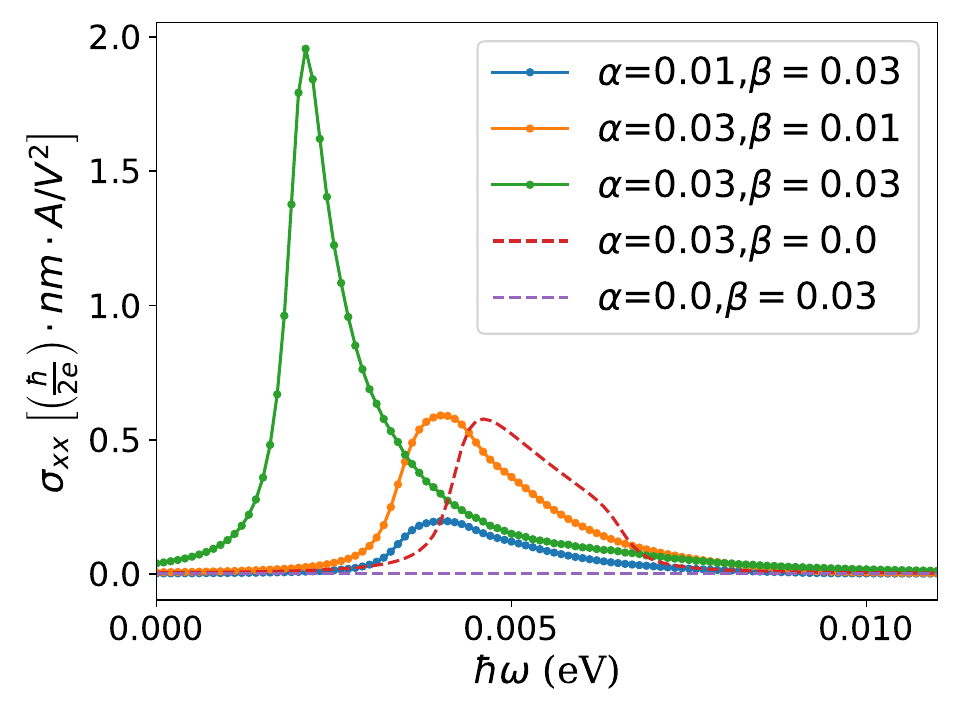}
	\caption{The longitudinal shift spin conductivity in the $k$-linear Rashba-Dresselhaus system with Zeeman coupling $\mu_z=0.001$ eV for the chemical potential at $\mu=0.005$ eV. The effective mass of an electron is taken to be $0.05\, m_0$, where $m_0$ is the free electron mass, in the calculation.  The transverse shift spin conductivity is similar and is shown in Fig. \ref{fig:lRDtrans} in the Appendix. Only the curves for $(\alpha,\beta)=(0.03, 0.01)$ and $(0.01, 0.03)$ are interchanged, apart from the overall sign difference. }
	\label{fig:lRD}
\end{figure}

\begin{figure}
	\includegraphics[width=0.45\textwidth]{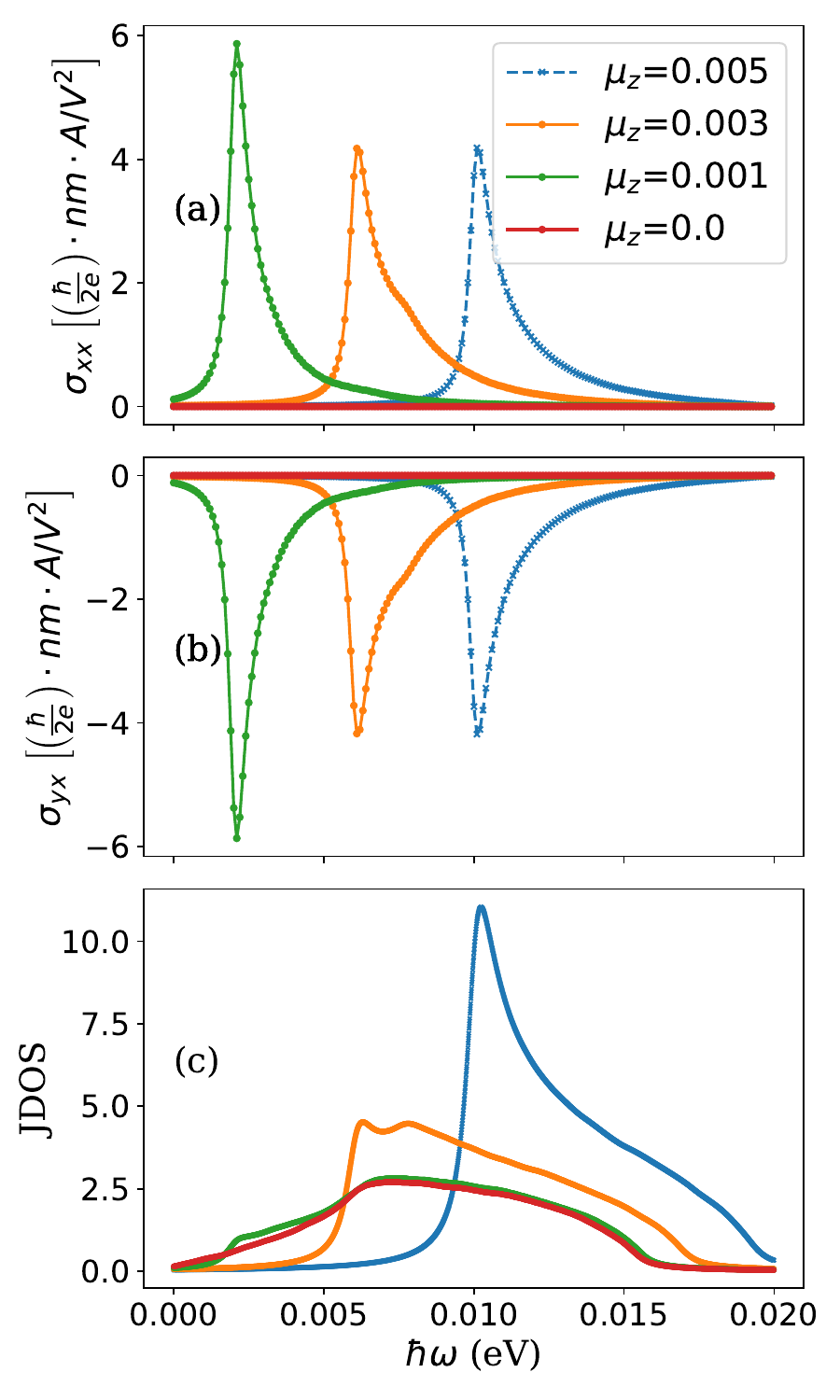}
	
	\caption{The longitudinal (a) and transverse (b) shift spin conductivity  and the joint density of states (c) for the $k$-linear Rashba-Dresselhaus system with various Zeeman coupling. $(\alpha, \beta)=(0.03, 0.03)$. The chemical potential is $\mu=0.005$ eV, within the energy gap. }
	\label{fig:lRDmuz}
\end{figure}

Consider the vanishing Zeeman coupling $\mu_z=0$. It can be shown that Eq. (\ref{Md-ML-3}) is still valid for the non-isotropic dispersion Eq. (\ref{NonIso-d}). Therefore, for any $k$-linear system of the dispersion $d=k\Gamma(\phi)$ with arbitrary function $\Gamma(\phi)$, including the $k$-linear Rashba-Dresselhaus system, the shift spin photocurrent always vanishes in the absence of Zeeman coupling. 
The details of the analytical calculation can be found in Appendix \ref{App:Matrix-dz} [see Eq. (\ref{App:M-Icab}) - Eq. (\ref{App:M-Icab2})]. %$G=\alpha_0^2-\beta_0^2$ [see Eq. (\ref{App:M-Icab2})], and we have $G=0$ for $\alpha_0=\beta_0$. 
 When there is non-zero Zeeman coupling ($\mu_z\neq0$), Eq. (\ref{Md-ML-1}) shows that the shift spin photocurrent could be nonvanishing, as shown in Fig. \ref{fig:lRD}.
  As shown in the figure, when $\alpha=\beta$, there is a strong response near $\hbar\omega=0.002$ eV. When $\alpha=\beta$, the Zeeman coupling destroys the degeneracy and opens up a finte gap at $\phi=0.25\pi, 1.25\pi$. If the chemical potential is within the gap, the joint density of states show van Hove singularies. The shift spin conductivities and the joint density of states for $\alpha=\beta$ with various Zeeman coupling are shown in Fig. \ref{fig:lRDmuz}. The shift spin conductivities (a,b) show strong peaks at $\hbar\omega=2\mu_z$, as a result of the van Hove singularities in the join density of states in (c). {The joint density of states increase as Zeeman coupling increases; however, the shift spin conductivies do not.} It suggests the dominant role of the matrix elements ($M^{Ic;ab}_{mn}$ in Eq. \ref{ShiftSpin-1}) in the shift spin conductivities.   

%\subsection{The cubic Rashba-Dresselhaus systems}

\begin{figure}
	\includegraphics[width=0.45\textwidth]{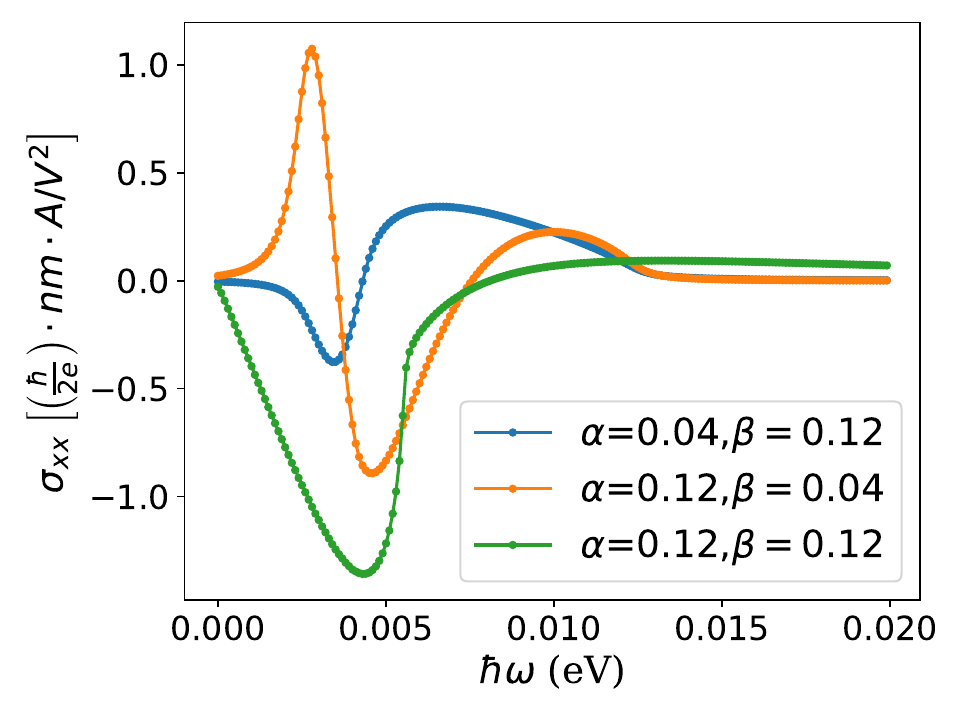}
	\caption{The longitudinal shift spin conductivity for the $k$-cubic Rashba-Dresselhaus system for chemical potential fixed at $\mu=0.01$ eV. The effective mass is set to be $0.27 m_0$ in the calculation.}
	\label{fig:cRD}
\end{figure}

\begin{figure}
		\includegraphics[width=0.45\textwidth]{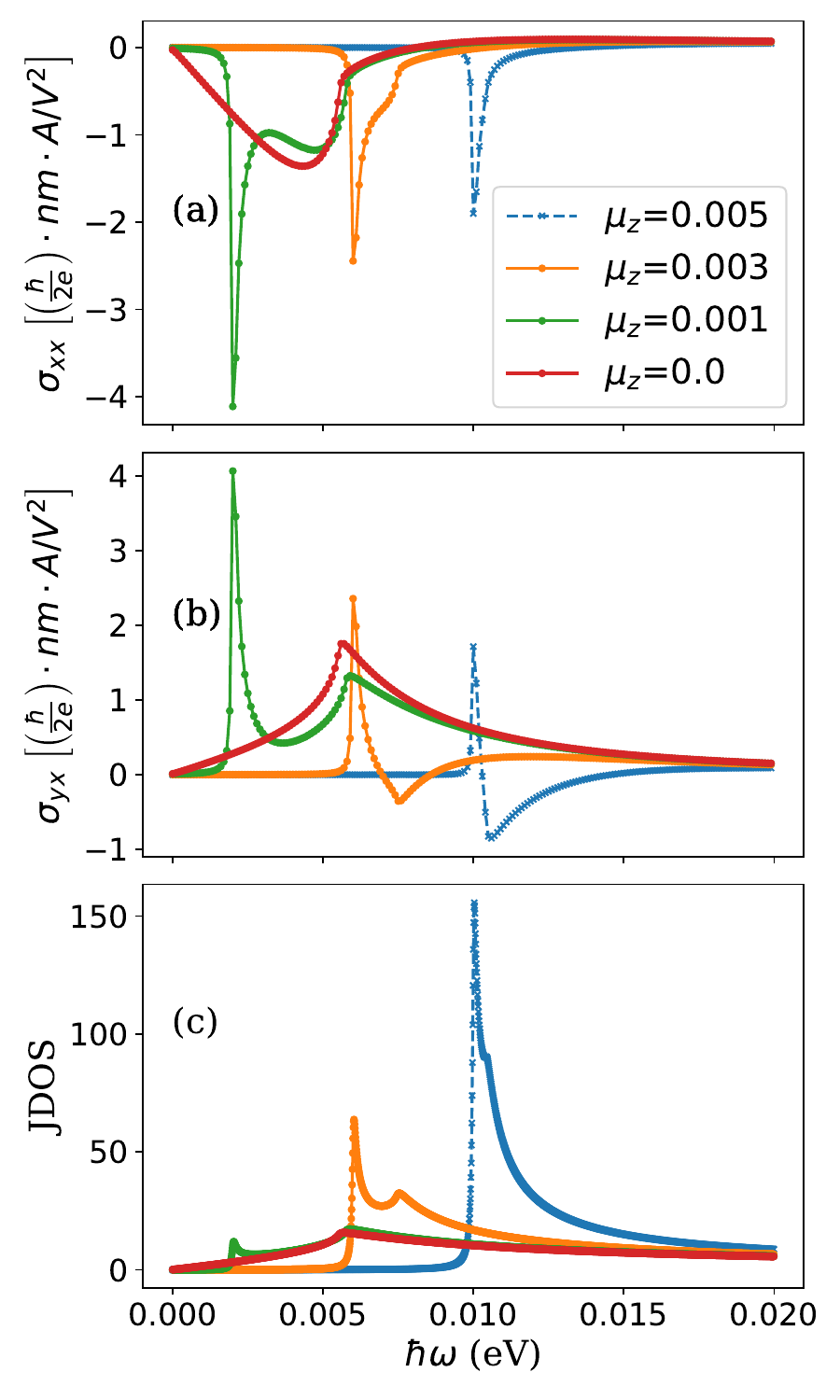}

	\caption{The longitudinal (a) and transverse (b) shift spin conductivities  and the joint density of states (c) for the $k$-cubic Rashba-Dresselhaus system with various Zeeman couplings. $(\alpha, \beta)=(0.12, 0.12)$. The chemical potential is $\mu=0.01$ eV, within the energy gap. }
	\label{fig:cRDmuz}
\end{figure}

For the $k$-cubic Rashba-Dresselhaus system, the longitudinal shift spin conductivities are shown in Fig. \ref{fig:cRD}. Unlike the $k$-linear Rashba-Dresselhaus systems, the shift spin photocurrent is nonvanishing even without Zeeman coupling. Like the $k$-linear Rashba-Dresselhaus systems, the parameters $\alpha=\beta$ also give rise to the strongest response. 
When a Zeeman coupling is introduced, the shift spin conductivities exhibit peaks, as shown in Fig. \ref{fig:cRDmuz} (a,b). Similar to the discussion in the previous paragraph for the $k$-linear Rashba-Dresselhaus system, 
the splitting of the degenerate energy bands leads to a singularity in the joint density of states, shown in Fig. \ref{fig:cRDmuz} (c). %leading to peaks in the shift spin conductivities. 

\begin{figure}
	\includegraphics[width=0.45\textwidth]{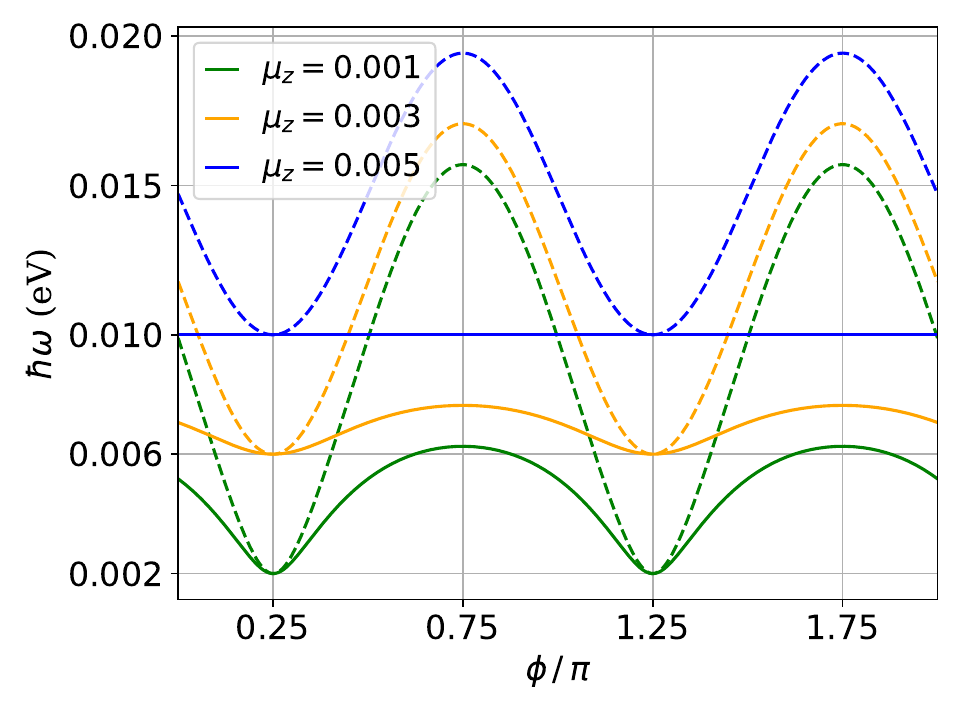}
	\includegraphics[width=0.45\textwidth]{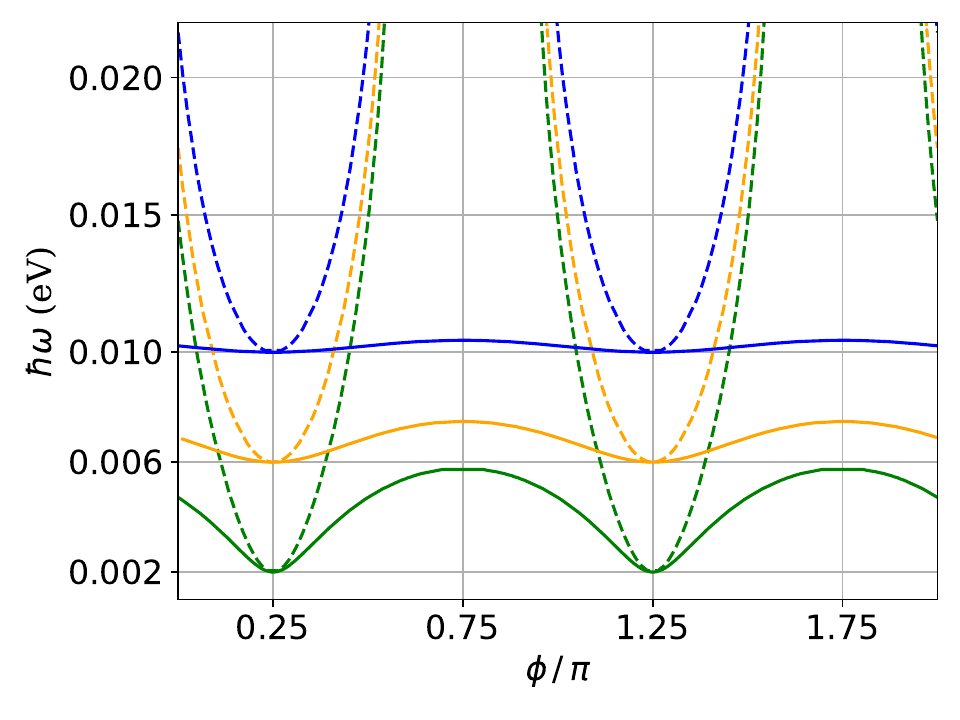}
	
	\caption{The maximum (dashed lines) and minimum (solid lines) allowed photon energies for the direct transition as a function of the azimuthal angle for the $k$-linear (top) and $k$-cubic (bottom) Rashba-Dresselhaus systems with $\alpha=\beta$ and various Zeeman couplings. }%For clarity of the figure, only $\mu_z=0.001$ and $\mu_z=0.005$ are shown. The curves for $\mu_z=0.001$ are in between.}
	\label{fig:lowedisp}
\end{figure}

Furthermore,  
\corr{the joint density of states (JDOS) is given by $\int d^2k \delta(E_--E_+-\hbar\omega)/(2\pi)^2$. For the integration in $\mathbf{k}$, the property for the Dirac $\delta$-function leads to $\delta(E_--E_+-\hbar\omega)=\frac{\delta(k-k(\omega))}{|\nabla_k(E_--E_+)|}$, where $k(\omega)=(E_--E_+)/\hbar$.  Thus, vanishing $|\nabla_k(E_--E_+)|$ leads to singularities in JDOS. } We plot the maximum and minimum allowed photon energy as a function of the azimuthal angle ($\phi$) in Fig. \ref{fig:lowedisp} and show that the van Hove singularities correspond to the region where $|\nabla_{{\bf k}}(E_--E_+)|$ is minimum.
There are two Fermi momenta in our two-band model, $k_f^n(\phi)$ with $n$ defined below Eq. (\ref{eq:eigen}). The maximum photon energy for the direct optical transition, denoted by $\Omega_{max}$, is given by $2d(k^-_f,\phi)$, whereas the minimum photon energy, denoted by $\Omega_{min}$, is given by $2d(k^+_f,\phi)$. As shown in Fig. \ref{fig:lowedisp}, for both $k$-linear and $k$-cubic Rashba-Dresselhaus systems, $\Omega_{min}=2\mu_z$ at $\phi=0.25\pi$ and $1.25\pi$. As the Zeeman coupling increases, $\Omega_{min}$ become more flat; thus $|\nabla_{{\bf k}}(E_--E_+)|$ decreases. This agrees with the join density of states shown in Fig. \ref{fig:lRDmuz} and \ref{fig:cRDmuz} (c).  

\subsection{Dirac surface states}
 The surface states of three-dimensional bismuth-chalcogenides-based topological insulators can be described by the effective Hamiltonian \cite{Fu2009,Ahn2020}
\begin{eqnarray}\label{eq:surfham}
	H_{D}=\epsilon_{\mk}+\hbar v(k_x\sigma_y-k_y\sigma_x)+\lambda(k_x^3-3k_xk_y^2)\sigma_z, 
\end{eqnarray}
where $v$ denotes the Dirac velocity and the last term is the hexagonal warping term accounting for the $C_{3v}$ lattice symmetry of the bulk Bi$_2$Te$_3$. The warping term was observed in angle-resolved photoemission spectroscopy (ARPES) \cite{Chen2009Exp} and explained by symmetry arguments by \cite{Fu2009}. The Hamiltonian preserves $M_x$ symmetry, where $M_x$ is the mirror reflection about the $yz$ plane,  $C_3$ rotation symmetry and  time-reversal symmetry $T$. The warping term breaks $M_y$ symmetry. 
%However, the combined $M_xT$ symmetry is still preserved.

% The charge photocurrent of surface states on the magnetic doped three-dimensional topological insualtor was observed in \cite{Ogawa2016} and theoretically discussed in \cite{Ahn2020}. 

The nonvanishing components of the shift spin conductivities can be determined by the symmetry analysis, as discussed in section \ref{sec:symm}. 
Because of the $M_x$ symmetry of the Hamiltonian in Eq. \ref{eq:surfham}, $v^x_{mn}$ is odd whereas $v^y_{mn}$ is even under this symmetry operation, explicitly $v^x_{mn}(-k_x,k_y)=-v^x_{mn}(k_x,k_y)$ and $v^y_{mn}(-k_x,k_y)=v^y_{mn}(k_x,k_y)$.
We consider circular polarized light incident perpendicularly to the surface; thus, $a=x, b=y$ in Eq. \ref{ShiftSpin-1}. For spin current moving in the $y$ direction, to have a nonvanishing conductivity, the spin component has to be $I=y,z$. In contrast, for spin current moving in the $x$ direction, the spin component for nonvanishing shift spin photocurrent is $I=x$.  

\begin{figure}
	\includegraphics[width=0.45\textwidth]{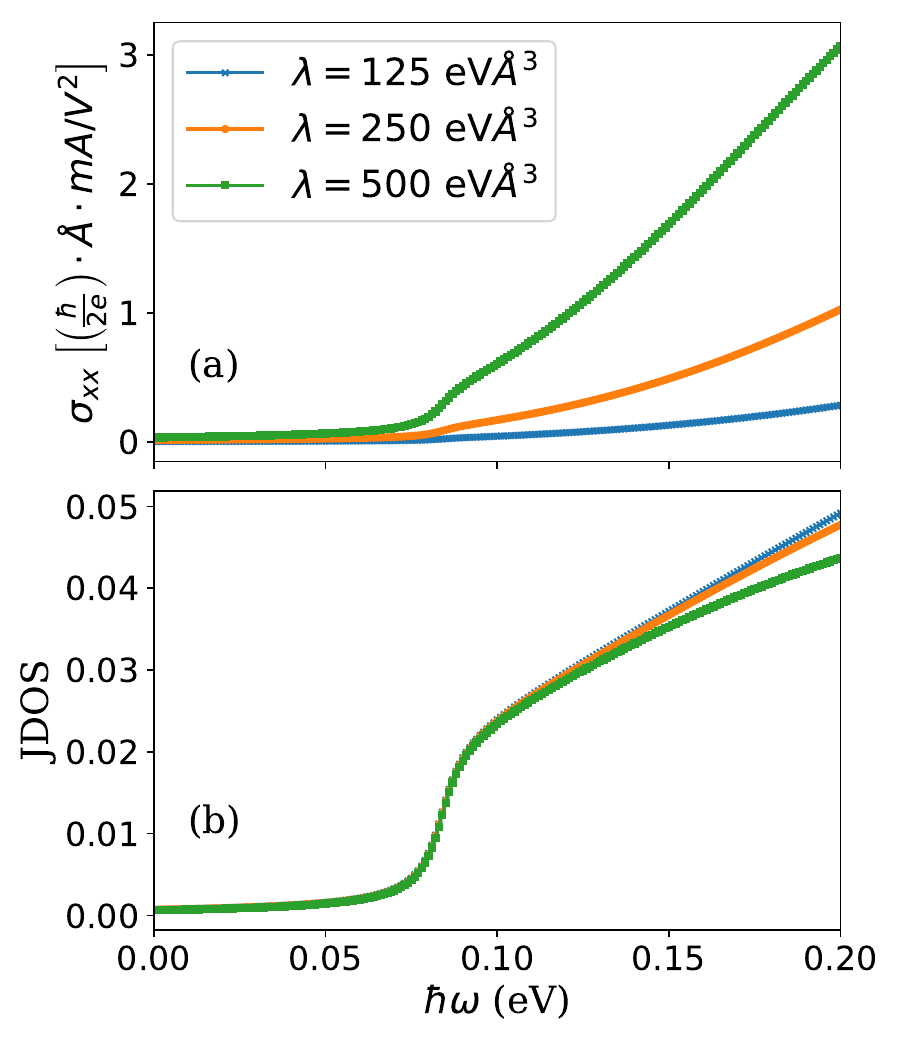}
	\caption{The longitudinal shift spin conductivity (a) and the joint density of states (b) for Dirac surface states on topological insulators.  The numerical values of the parameters for the calculations are $m=0.13 m_0, \hbar v=2.5$ eV$\AA$, and the chemical potential is set at $0.05$ eV.}
	\label{fig:yy}
\end{figure}

{
The numerical results of the shift spin conductivity for the effective Hamiltonian Eq. \ref{eq:surfham} with different warping strengths are shown in Fig. \ref{fig:yy}. The shift spin conductivity grows as the warping term becomes stronger. In contrast, the joint density of states decrease mildly only at higher photon frequency. It shows that the warping term has a significant impact on the matrix element Re$\left[M_{nm}^{xx,xy}\right]$.} 
%The plot of the integrand %Re$\left[M_{nm}^{xx,xy}\right]$ in momentum space is given in Fig. \ref{fig:yycont}, showing sixfold symmetry, the same symmetry as the energy levels. This is a consequence of the $C_3$ symmetry and time-reversal symmetry of the Hamiltonian. 
All components have been calculated and only the significant nonvanishing component, $\sigma^{xx,xy}$, is shown. $\sigma^{zy,xy}$ is one order of magnitude smaller and not shown. 
The two longitudinal circurlar shift spin photocurrents are equal, i.e.  $\sigma^{yy,xy}=\sigma^{xx,xy}$. Thus, we only show $\sigma^{xx,xy}$. 

Compared to the study of the spin photocurrent of the Dirac surface states in \cite{Kim2017}, where time-reversal symmetry is broken, in our calculation, the shift spin photocurrent is generated without breaking time-reversal symmetry and the shift charge current is zero. Nonetheless, we include the warping term, which is higher-order in $k$, that can lead to nonvanishing shift spin photocurrent.

\section{Conclusion}\label{sec:disc}
In this paper, we have theoreticaly shown that in two-dimensional systems, the shift spin photocurrent can be generated by circularly polarized light. %, dubbed as the shift spin photocurrent. The typical two-dimensional spin-orbit coupled systems are studied, the $k$-linear and $k$-cubic Rashba-Dresselhaus systems, Wurtzite Hamiltonian and Dirac surface states. 
For the $k$-linear Rashba-Dresselhaus systems, shift spin photocurrent always vanishes, unless a Zeeman energy is coupled in the system. For the $k$-cubic Rashba-Dresselhaus systems and Wurtzite systems, the shift spin photocurrents are nonzero. When Zeeman coupling is introduced, the band-splitting leads to van Hove singularities when the states for allowed direct transitions are near the band bottom, creating peaks of the shift spin photocurrent at corresponding frequencies. %It worths noting that merely increasing the van hove singularity does not strength the shift spin photocurrent. The matrix element of the shift conductivity 

Furthermore, by symmetry analysis, we show that in Rashba type systems where mirror symmetries $M_x$ and $M_y$ are preserved, the spin polarization and the spin current directions are parallel. Longitudinal components of shift spin conductivities would survive{, while the transverse components always vanish.} In Dresselhaus type systems where the parity-mirror symmetry is preserved, the spin polarization and the spin current directions are perpendicular. In this case, transverse components of shift spin conductivities would survive{, while the longitudinal components always vanish.} For Dirac surface states with warping term, because of the remaining $M_x$ symmetry, the longitudinal shift spin photocurrent is nonvanishing and increases with the strength of the warping term. The results in this work could be applied for the design of spintronic devices for spin current with particular spin polarization. 

%\corr{Before concluding, we would like to point out that recent studies have shown that altermagnets can support pure spin current regardless of SOC \cite{Dong2024,Pan2024}. However, the shift spin photocurrent driven by circularly polarized light is zero in altermagnetics when SOC is absent, as idicated by the symmetry analysis in \cite{Dong2024}.   } 
\begin{acknowledgments}
	H.-C. H. would like to thank Prof. Tay-Rong Chang and Dr. Yang-hao Chan for insightful discussions. 
	H.-C. H acknowledges the support from the National Science and Technology Council (NSTC) and the National Center for Theoretical Sciences (NCTS) in Taiwan. H.-C. H. was supported by NSTC under Grant No. 113-2628-M-004-001-MY3.
\end{acknowledgments}

\appendix
\comm{\section{Spin operator in $k$-cubic Hamiltonian}\label{app:spinop}}
\comm{In this Appendix, we address the problems in the definition of the effective spin operator in $k$-cubic spin-orbit coupled systems. The Hamiltonian of the hole gas is given by the sum of both Luttinger-Kohn Hamiltonian $H_{LK}$~\cite{Kohn1955,Luttinger1956} and structure inversion asymmetric (SIA) Hamiltonian $H_R$ and bulk inversion asymmetric (BIA) Hamiltonian $H_D$,
\begin{equation}\label{App:LH1}
	H=H_{LK}+H_R+H_D+V_c(z),
\end{equation}
where $V_c(z)$ is the confinement potential. The SIA Hamiltonian~\cite{Winkler2000-1,Winkler2000-2} is given by
\begin{equation}
	H_R=\alpha_1(\mathbf{k}\times\mathbf{J})\cdot\hat{e}_z
\end{equation}
and the BIA Hamiltonian~\cite{Dress1955-1} is given by
\begin{equation}
	H_D=\beta_1\mathbf{J}\cdot\boldsymbol{\Omega},
\end{equation}
where $\Omega_z=k_z(k_x^2-k_y^2)$, $\Omega_x=k_x(k_y^2-k_z^2)$ and $\Omega_y=k_y(k_z^2-k_x^2)$.}

\comm{The Hamiltonian Eq. (\ref{App:LH1}) is written in the $|J,J_z\rangle$ basis ($|3/2,3/2\rangle,|3/2,1/2\rangle,|3/2,{-1/2}\rangle,|3/2,1/2\rangle$), describing the light-hole ($|3/2,{\pm1/2}\rangle$) band and heavy hole bands ($|3/2,{\pm3/2}\rangle$). The matrix $J_z$ in this basis is written as
\begin{equation}
	J_z=\left(\begin{array}{cccc}
		3/2&0&0&0\\
		0&1/2&0&0\\
		0&0&-1/2&0\\
		0&0&0&-3/2\\
	\end{array}\right).
\end{equation}
The $x$ and $y$ components are given by
\begin{equation}
	J_x=\left(\begin{array}{cccc}
		0&\sqrt{3}/2&0&0\\
		\sqrt{3}/2&0&1&0\\
		0&1&0&\sqrt{3}/2\\
		0&0&\sqrt{3}/2&0\\
	\end{array}\right),
\end{equation}
and
\begin{equation}
	J_y=\left(\begin{array}{cccc}
		0&-i\sqrt{3}/2&0&0\\
		i\sqrt{3}/2&0&-i&0\\
		0&i&0&-i\sqrt{3}/2\\
		0&0&i\sqrt{3}/2&0\\
	\end{array}\right).
\end{equation}
Thus, the Hamiltonian Eq. (\ref{App:LH1}) is a $4\times4$ matrix. If only the heavy-hole band is concerned, we have to use the L$\ddot{o}$wdin perturbation method~\cite{Pikus1974,Winkler} to project the Hamiltonian Eq. (\ref{App:LH1}) into a subspace spanned by $|3/2,{\pm3/2}\rangle$. The resulting Hamiltonian $H^{hh}_{2D}$ describing the heavy hole subspace is then a $2\times2$ matrix composed of the sum of $k$-cubic Rashba Hamiltonian $H^{hh}_R$ and $k$-cubic Dresselhaus Hamiltonian $H^{hh}_D$ shown in Ref.~\cite{Loss2005}, where
\begin{equation}\label{App:k3R}
	H^{hh}_R=i\alpha(k_-^3\sigma_+-k_+^3\sigma_-)
\end{equation}
and
%is the Rashba Hamiltonian and
\begin{equation}\label{App:k3D}
	H^{hh}_D=-\beta(\sigma_+k_+k_-^2+\sigma_-k_-k_+^2).
\end{equation}
%is the Dresselhaus Hamiltonian.
The spin-orbit parameters $\alpha$ and $\beta$ in Eqs. (\ref{App:k3R}) and (\ref{App:k3D}) has been theoretically and experimentally studied~\cite{Winkler2000-2,Loss2005,Moriya2014}.}

\comm{The Pauli matrices in Hamiltonian $H^{hh}_R$ and $H^{hh}_D$ can be regarded as pseudospins. We identify the effective spin eigenstates with the $J_z = \pm3/2$ eigenvalues, i.e. $|{\uparrow}\rangle=|3/2,{+3/2}\rangle$ and $|{\downarrow}\rangle=|3/2,{-3/2}\rangle$. We assume spin $3\hbar/2$ for the heavy holes, and thus, the effective spin operator for the hole particle should be given by
\begin{equation}\label{App:spin}
	\mathbf{s}=\frac{3}{2}\hbar\boldsymbol{\sigma}
\end{equation}
for three spin components and the factor of $3/2$ reflects the total angular momentum quantum numbers of the heavy holes. We stress that the overall factor $3/2$ does not affect the results if in a steady state the direction of spins is assumed to be parallel to the spin-orbit field $\mathbf{d}(\mathbf{k})/d$, as used in Ref.~\cite{Wund2005} for studying the dynamics of the direction of spins in the 2D hole system.}

%\comm{{\bf This paragraph is a bit confusing:} On the other hand, another issue arises: the spin operator would also be defined by the L$\ddot{o}$wdin perturbation method which results in the hybridization of spin and orbit angular momentum. In Ref~\cite{Shen2008}, the eight-band model Hamiltonian was used, and the resulting effective in-plane spin operators depends on the momentum $\mathbf{k}$ (see Eq. (29) and (30) in~\cite{Shen2008}), and the effective spin-z component is still $s_z=(3/2)\hbar\sigma_z$. The k-dependent spin operator in Ref.~\cite{Shen2008} was used in Ref.~\cite{Bar2018} for studying the non-equilibrium spin polarization induced by an external electric field.}

\comm{In this paper, we use Eq. (\ref{App:spin}) as the effective spin operator because the effect of cubic Rashba spin splitting Eq. (\ref{App:k3R}) not only appears in 2D hole system, but also in the quasi-2D electron gas~\cite{Nakamura2012,Zhao2020}. Furthermore, the use of Eq. (\ref{App:spin}) leads to the non-zero and zero shift spin currents in the $k$-cubic Dresselhaus and Rashba systems, respectively. The important difference between the $k$-cubic Rashba and Dresselhaus systems and their interplay in the shift spin current should be determined by experiment in the future.}
{\section{Analytical derivations}}
\subsection{Matrix elements of spin}
\label{App:Spin}
The diagonal matrix elements of Pauli matrices in the basis of Eq. (\ref{H1-vector}) are given by
\begin{equation}\label{Pauli-1}
	\begin{split}
		&\langle{n\mk}|\sigma_x|{n\mk}\rangle=-n\hat{d}_x,\\
		&\langle{n\mk}|\sigma_y|{n\mk}\rangle=-n\hat{d}_y,\\
		&\langle{n\mk}|\sigma_z|{n\mk}\rangle=-n\hat{d}_z,\\
	\end{split}
\end{equation}
The off-diagonal matrix elements of Pauli matrices are given by
\begin{equation}\label{Pauli-2}
	\begin{split}
		&\langle{+\mk}|\sigma_x|{-\mk}\rangle=\hat{d}_z+\frac{(-i)\hat{d}_y(\hat{d}_x+i\hat{d}_y)}{1+\hat{d}_z},\\
		&\langle{+\mk}|\sigma_y|{-\mk}\rangle=i\hat{d}_z+\frac{(+i)\hat{d}_x(\hat{d}_x+i\hat{d}_y)}{1+\hat{d}_z},\\
		&\langle{+\mk}|\sigma_z|{-\mk}\rangle=-(\hat{d}_x+i\hat{d}_y),\\
	\end{split}
\end{equation}
where $\hat{d}_x^2+\hat{d}_y^2+\hat{d}_z^2=1$ was used. We also note that Eqs. (\ref{Pauli-1}) and (\ref{Pauli-2}) also satisfy the matrix elements of the commutator $[\sigma_i,H_0]=2i\epsilon_{ijk}d_j\sigma_k$.

\subsection{Matrix elements $M^{Ic,ab}$ for $d_z\neq 0$}\label{App:Matrix-dz}
Suppose that $d_z=d_z(\mk)$, we have
\begin{widetext}
	\begin{equation}\label{App:M-Icab}
		\begin{split}
			&Re[M^{yy,xy}_{-1,1}]=N\left\{(-\hat{d}_z^2)\left[\frac{\partial d_y}{\partial k_x}\left(1-\frac{k_y}{d}\frac{\partial d}{\partial k_y}\right)+\frac{\partial d_y}{\partial k_y}\left(\frac{k_y}{d}\frac{\partial d}{\partial k_x}\right)\right]+G\frac{\hat{d}_x}{d}k_y\left(1-\frac{k}{d}\frac{\partial d}{\partial k}\right)\right\},\\
			&Re[M^{xy,xy}_{-1,1}]=N\left\{(-\hat{d}_z^2)\left[\frac{\partial d_x}{\partial k_x}\left(1-\frac{k_y}{d}\frac{\partial d}{\partial k_y}\right)+\frac{\partial d_x}{\partial k_y}\left(\frac{k_y}{d}\frac{\partial d}{\partial k_x}\right)\right]-G\frac{\hat{d}_y}{d}k_y\left(1-\frac{k}{d}\frac{\partial d}{\partial k}\right)\right\},\\
			&Re[M^{xx,xy}_{-1,1}]=N\left\{(+\hat{d}_z^2)\left[\frac{\partial d_x}{\partial k_y}\left(1-\frac{k_x}{d}\frac{\partial d}{\partial k_x}\right)+\frac{\partial d_x}{\partial k_x}\left(\frac{k_x}{d}\frac{\partial d}{\partial k_y}\right)\right]-G\frac{\hat{d}_y}{d}k_x\left(1-\frac{k}{d}\frac{\partial d}{\partial k}\right)\right\},\\
			&Re[M^{yx,xy}_{-1,1}]=N\left\{(+\hat{d}_z^2)\left[\frac{\partial d_y}{\partial k_y}\left(1-\frac{k_x}{d}\frac{\partial d}{\partial k_x}\right)+\frac{\partial d_y}{\partial k_x}\left(\frac{k_x}{d}\frac{\partial d}{\partial k_y}\right)\right]+G\frac{\hat{d}_x}{d}k_x\left(1-\frac{k}{d}\frac{\partial d}{\partial k}\right)\right\},\\
			&Re[M^{zx,xy}_{-1,1}]=N\hat{d}_z\left[-\hat{d}_z\frac{k_x}{d}\left(\frac{\partial d_z}{\partial k_y}\frac{\partial d}{\partial k_x}-\frac{\partial d_z}{\partial k_x}\frac{\partial d}{\partial k_y}\right)-\frac{\partial d}{\partial k_y}+\hat{d}_z\frac{\partial d_z}{\partial k_y}\right],\\
			&Re[M^{zy,xy}_{-1,1}]=N\hat{d}_z\left[\hat{d}_z\frac{k_y}{d}\left(\frac{\partial d_z}{\partial k_x}\frac{\partial d}{\partial k_y}-\frac{\partial d_z}{\partial k_y}\frac{\partial d}{\partial k_x}\right)+\frac{\partial d}{\partial k_x}-\hat{d}_z\frac{\partial d_z}{\partial k_x}\right],\\
		\end{split}
	\end{equation}
\end{widetext}
where the term $G$ is defined as
\begin{equation}\label{App:Def-G}
	\begin{split}
		&d_x\frac{\partial d_y}{\partial k_x}-d_y\frac{\partial d_x}{\partial k_x}=-Gk_y,\\
		&d_x\frac{\partial d_y}{\partial k_y}-d_y\frac{\partial d_x}{\partial k_y}=+Gk_x,
	\end{split}
\end{equation}
and for the systems under consideration we have
\begin{equation}\label{App:M-Icab2}
	G=\det(\tilde{\beta})=\beta_{xx}\beta_{yy}-\beta_{yx}\beta_{xy}
\end{equation}
for k-linear system. By using Table \ref{Table-SOC-k1}, we have $\det(\tilde{\beta})=\alpha_0^2-\beta_0^2$ for Rashba-Dresselhaus system,  $\det(\tilde{\beta})=\eta^2$ for Weyl system and $G=0$ for PST system. For k-cubic Rashba and Dresselhaus system, we have
\begin{equation}\label{App:M-Icab3}
	G=G(k,\phi)=(3\alpha^2+\beta^2)k^4-8\alpha\beta k^4\cos\phi\sin\phi.
\end{equation}
For Wurtize system, it is given by
\begin{equation}
	G=(\alpha+\beta k^2)^2.
\end{equation}
The overall factor $N$ is given by
\begin{equation}\label{App:M-Icab4}
	N=\frac{\hbar}{4d^2}\frac{q\hbar^2}{2m},
\end{equation}
where $q=1$ for k-linear system and $q=3$ for k-cubic Rashba and Dresselhaus system.

\section{Additional numerical results}\label{app:numer}
\begin{figure}[h!]
	\includegraphics[width=0.5\textwidth]{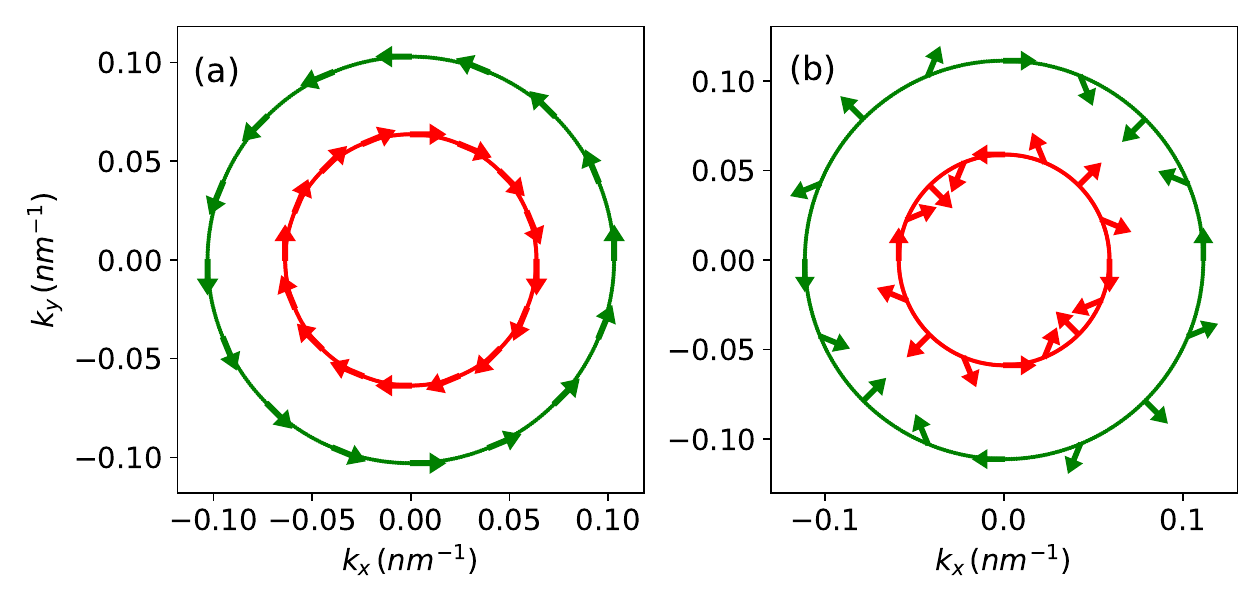}
	\caption{\comm{The schematic of the spin textures on the Fermi contours for the $k$-linear Rashba (a) and the $k$-cubic Rashba (b) Hamiltonian.}}
	\label{fig:spintext}
\end{figure}

\comm{The spin textures of the $k$-linear and cubic Hamiltonians are given by Eq. \ref{Pauli-1} and shown in Fig. \ref{fig:spintext}.}

\begin{figure}[h!]
	\includegraphics[width=0.45\textwidth]{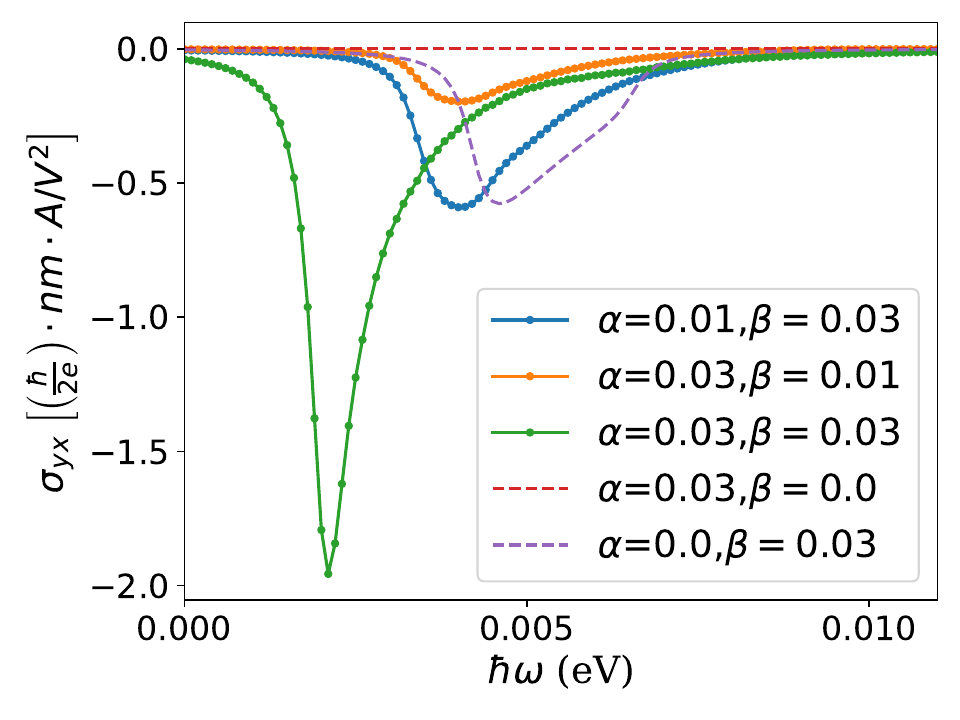}
	\caption{The transverse shift spin conductivity for the  $k$-linear Rashba-Dresselhaus system with out-of-plane Zeeman $\mu_z=0.001$ eV. Compared to the longitudinal shift spin conductivity in Fig. \ref{fig:lRD}, only the curves for different $\alpha,\beta$ are switched apart from the overall sign difference. The parameters for the calculation are the same as in Fig. \ref{fig:lRD}. }
	\label{fig:lRDtrans}
\end{figure}

The transverse shift spin conductivity for the linear Rashba-Dresselhaus is shown in Fig. \ref{fig:lRDtrans}. 

%The longitudinal and transverse shift spin conductivities and joint density of states for the persistent-spin-texture (PST) model are shown in Fig. \ref{fig:PSTmuz}
% \begin{figure}
% 	\includegraphics[width=0.45\textwidth]{./pst_jdos_muz.pdf}	
% 	\caption{The longitudinal (a) and transverse (b) shift spin conductivity  and the joint density of states (c) for the persistent-spin-texture (PST) model with various Zeeman coupling. The spin-orbit coupling strength ($\lambda$) is $0.03$ eV$\cdot$nm in the calculation. The chemical potential is $\mu=0.005$ eV, within the energy gap.}	\label{fig:PSTmuz}
 %\end{figure} 
%In the main text, we discuss the impact of the van Hove singularities on the shift spin conductivity. The dispersions across a given Fermi line are shown in Fig. \ref{fig:bands}. It shows that the allowed transition regions are near the band bottoms. 
%\begin{figure}	\includegraphics[width=\linewidth]{./bands.pdf}
%	\caption{ The energy bands near the allowed transition region for the $k$-linear (a) $k$-cubic (b). The model paramters for the $k$-linear Hamiltonian are $(\alpha, \beta)=(0.03, 0.03)$ eV$\cdot$ nm, $\mu_z=0.001$ eV, $\mu=0.005$ eV and for the $k$-cubic Hamiltonian are $(\alpha, \beta)=(0.12, 0.12)$ eV$\cdot$ nm$^{-3}$, $\mu_z=0.001$ eV, $\mu=0.01$ eV. The orange dashed lines denote the energy levels that the Fermi electrons transition to when $\hbar\omega=0.002$ eV in both figures.  The black dashed lines denote chemical potential. }	\label{fig:bands}
%\end{figure} 

\bibliography{references,referencesInpaper}      

\end{document}